\documentclass[fleqn, onecolumn, usenatbib, useAMS]{mnras}

\usepackage{graphicx}	
\usepackage{amsmath}	
\usepackage{amssymb}	
\usepackage{multicol}        

\usepackage{bm}		
\usepackage{pdflscape}	

\def\bea{\begin{eqnarray}}
\def\eea{\end{eqnarray}}
\def\ba{\begin{array}}
	\def\ea{\end{array}}

\def\beq{\begin{equation}}
\def\eeq{\end{equation}}

\newcommand{\cl}{\ell}

\newcommand{\eq}{Eq.\eqref}
\newcommand{\fig}{Fig.\ref}

\usepackage[T1]{fontenc}
\usepackage{ae,aecompl}

\usepackage{times}

\title[Reconstruction of  Intrinsic $B$-mode Power]{\textsc{A Demonstration of Spectral Level Reconstruction of  Intrinsic $B$-mode Power}}

\author[Barun Pal]{Barun  Pal\thanks{Contact e-mail: \href{terminatorbarun@gmail.com}{terminatorbarun@gmail.com}}
\\
Department of Mathematics, Netaji Nagar College for Women, Kolkata -- 700092, India}

\pubyear{2016}

\begin{document}
\label{firstpage}
\pagerange{\pageref{firstpage}--\pageref{lastpage}}
\maketitle

\begin{abstract}
We investigate the prospects and consequences of the spectral level reconstruction of primordial $B$-mode power by solving the systems of linear equations assuming that the lensing potential together with the lensed polarization spectra are already in hand. We find that this reconstruction technique may be very useful to  have an estimate of the amplitude of primordial gravity waves or more specifically the value of tensor to scalar ratio. We also see that one can have cosmic variance limited reconstruction of the intrinsic $B$-mode power up to few hundred multipoles ($\cl\sim500$) which is more than sufficient to have an estimate of the tensor to scalar ratio. Since the small scale cosmic microwave background (CMB henceforth) anisotropies  are not sourced by the primordial gravity waves generated during inflation. 
We also find that the impact of instrumental noise  may be bypassed within this reconstruction algorithm. A simple demonstration for the nullification of the instrumental noise anticipating \textit{COrE} like futuristic space mission complemented with \textit{Planck} 2013 cosmology  has been presented. 
\end{abstract}




\begingroup
\let\clearpage\relax
\endgroup

\section{Introduction}
The existence of gravity waves was predicted long ago by none other than \textit{Albert Einstein}. After a century's intermission finally we have detected the gravity waves first from a binary black hole merger \citep{ligo2016a} and later from a 22-Solar-Mass binary black hole coalescence \citep{ligo2016b} jointly by LIGO\footnote{\url{https://www.ligo.caltech.edu/}} and Virgo\footnote{\url{http://www.virgo-gw.eu/}} scientific collaborations. As a consequence  people are now eagerly waiting for the primordial gravity waves which are believed to be produced  during cosmic inflation \citep{starobinsky1979, guth1981} through the tensor perturbations along with the primordial density perturbations.  As the size of co-moving horizon decreases during inflation, all the modes leave the Hubble radius. The tensor modes remain constant after horizon exit just like the scalar modes \citep{lyth2005}. Once inflation is over the co-moving size of the horizon starts growing again and eventually all the modes reenter Hubble radius. Tensor modes with smaller wavelengths make horizon reentry early and decay very quickly ahead of recombination. As a result small scale CMB anisotropies are not affected by the  tensor perturbations, only the large scale anisotropies get contribution from the gravity waves. The large scale CMB $B$-mode polarization are believed to be sourced solely by the primordial gravity waves \citep{seljak1997}. Accordingly, the intrinsic $B$-mode has very little power on the smaller scale ($\cl\geq 200$) and we only concentrate on the large scale CMB polarizations for primordial gravity waves.

 The measurement of large scale $B$-mode amplitude will help us determine the energy scale of inflation as it is directly related to the primordial gravity waves. But during their journey from last scattering surface to the present day detectors  CMB photons encounter several over and under dense regions which perturbs their paths -- a phenomena known as gravitational lensing of CMB \citep{linder1990, seljak1996, metcalf1997,zaldarriaga1998, lewis2006}. As a result the CMB power spectra are modified which is the worst in case of $B$-mode. So the subtraction of lensing contribution from the $B$-mode signal is more than a necessary. The large scale CMB $B$- mode is the primary source of  primordial gravity waves. But the transfer of power from larger to smaller scales due to gravitational lensing eludes intrinsic CMB \citep{zaldarriaga1998, seljak2004}. Subtraction of lensing confusion from the large scale $B$-mode will certainly improve our current understanding about cosmic inflation \citep{knox2002, simard2015}. So, in order to get rid of the lensing artifacts delensing is a bare necessity. 

 There are many delensing techniques   available in the literature. Most of them rely on the reconstruction of lensing potential \citep{hu2003, kesden2003, hirata2003, hanson2011, namikawa2012, namikawa2014, pearson2014}. Some of them utilize external data set for delensing \citep{sigurdson2005, smith2012, sherwin2015}. Very recently, cosmic infrared background has been utilized to reverse the gravitational lensing of CMB \citep{larsen2016}.  In this work we shall be discussing about the prospects and consequences of the reconstruction of intrinsic $B$-mode power through direct matrix inversion technique \citep{barun2014}, assuming the availability of lensing potential and lensed polarization spectra. In this process we first evaluate the lensing kernels for polarization spectra and invert them applying {\it Gauss-Jordan} elimination technique \citep{press2007}. With the help of the inverted kernel matrices we solve for intrinsic CMB polarization spectra. In order to estimate the error in the reconstructed intrinsic CMB $B$-mode power we have made use of simulation. Initially thousand realizations of lensed CMB for four different values of tensor to scalar ratios, $r=0, \ 0.01, \ 0.005 \ \mbox{and} \ 0.001$, have been generated keeping the lensing potential fixed. We then apply our delensing algorithm to obtain the realizations of the delensed  $C_\cl^B$. The mean and the error in the reconstructed $B$-mode power are then calculated from those delensed realizations. To get the actual power of the intrinsic $B$-mode we need to subtract the delensing bias and we find that bias subtracted mean of the recovered $B$-mode power matches very well with the injected $C_\cl^B$. We also find that one can have cosmic variance limited (CVL henceforth) reconstruction of intrinsic $C_\ell^B$ almost up to $\cl\sim500$ even for tensor to scalar ratio three orders of magnitude below unity. Though we are far away to detect  primordial gravity waves for $r=0.001$, but the futuristic space mission in the likes of  PRISM\footnote{\url{http://www.prism-mission.org/}}, COrE\footnote{\url{http://www.core-mission.org/science.php}} are promising to detect gravity waves with tensor to scalar ratio as low as $10^{-3}$ or even a order lower.  We have also found that the incorporation of instrumental noise anticipating COrE  like futuristic space mission has almost null impact on our reconstruction procedure.  In fact the effect of any instrumental noise may be nullified within our delensing algorithm. But, we do not make definitive remarks on this as the non-Gaussian structure of the lensed CMB,  the lensing by gravity waves and the errors on the cosmological parameters have been neglected along with fact that current observations covers only $70-80\%$ of the sky.  
 
 For our entire analysis we have used the following {\it Planck + WMAP} best fit $\Lambda$CDM cosmology  \citep{planck2013}: 
 $\Omega_bh^2=0.022032$, $\Omega_ch^2=0.12038$, $\Omega_\lambda=0.6817$, $H_0=67.04\ {\rm Km/Sec/Mpc}$, $n_{_S}=0.9619$, $P_{\cal R}=2.215\times 10^{-9}$  and $\tau=0.0925$. We have used publicly available code CAMB\footnote{\url{http://camb.info/}} to compute the theoretical CMB lensing power spectrum as well as lensed spectra for the above cosmology. In our analysis we did not take into account lensing due to the gravity waves as they are expected to be very small. For the instrumental noise we have considered 225 GHz frequency channel of the upcoming satellite mission COrE.
 
 The paper is organized as follows. In Section \ref{mit} we have reviewed the delensing through direct matrix inversion technique. In section \ref{bias} we have presented our findings for a single realization of lensed CMB along with the delensing bias and results for simulation in the absence of instrumental noise.  In Section \ref{del_noise} we have discussed prospect of $B$-mode delensing in presence of COrE like instrumental noise. We summarize our findings and discuss future prospects in a brief concluding section. 
 
\section{Brief Review of Delensing Through Direct Matrix Inversion}\label{mit}
The path of the CMB photon is perturbed due to the gravitational lensing by the potential gradients transverse to the line of sight. 
As a result a point $\hat{\bf n}$, on the last scattering surface (LSS) appears to be in a deflected position $\hat{\bf n}'$. 
The (lensed) temperature $\tilde{T}(\bf \hat{n})$ that we measure as coming from a direction $\bf \hat{n}$ in the sky, actually 
corresponds to the intrinsic temperature ${T}(\bf \hat{n}')$ from a different direction $\bf \hat{n}'$, where $\bf \hat{n}$ and $\bf \hat{n}'$ are 
related through the deflection angle $\bm{\alpha}$ with $\bf \hat{n}' = \bf \hat{n} + \bm{\alpha}$. 
Similarly, the polarization field is remapped according to $\tilde{\bf P}(\hat{\bf n})={\bf P}(\hat{\bf n}')$. 

\subsection{Lensed CMB Power Spectra}
The lensed temperature power spectrum in the full spherical sky limit using correlation function technique can be written as \citep{seljak1996, lewis2005}
\beq\label{lensed_clt}
\tilde{C}_\ell^{T}= \sum_{\ell'} K^T_{\ell \ell'} C_{\ell'}^{T}
\eeq
where $K^T_{\ell \ell'}$ is  lensing kernel associated with the temperature field given by the following expression
\bea
K^T_{\ell \ell'} &=&\delta_{\cl\cl'}+\frac{2\ell'+1}{2}\int^\pi_{0} \sin \beta \ d \beta \ d^\ell_{00}(\beta)\left\{\left[X^2_{000}-1\right]d^{\ell'}_{00}(\beta)+
\frac{8}{\ell'(\ell'+1)} A_2 (\beta) X'^2_{000}d^{\ell'}_{1 -1}(\beta) \right.\nonumber\\
&+&\left.
A_2 (\beta) ^2\left(X'^2_{000}d^{\ell'}_{00}(\beta)+X^2_{220}d^{\ell'}_{2-2}(\beta)\right)\right\}.
\eea
Here $\cos\beta = \hat{\bf n}_1 \cdot \hat{\bf n}_2$, $d^\ell_{mm'}$'s are the standard Wigner rotation matrices and  we have defined,
\bea
A_0(\beta)&\equiv&\sum_{\ell}\frac{2\ell+1}{4\pi}\ell(\ell+1)C_\ell^{\phi}d^\ell_{11}(\beta) ;\
A_2 (\beta)\equiv\sum_{\ell}\frac{2\ell+1}{4\pi}\ell(\ell+1)C_\ell^{\phi}d^\ell_{-11}(\beta);  \
\nonumber\\
\sigma^2(\beta)&\equiv&A_0(0)-A_0(\beta); \ X_{imn}\equiv \int_0^{\infty}\frac{2\alpha}{\sigma^2(\beta)}\left(\frac{\alpha}{\sigma^2(\beta)}\right)^i e^{-\alpha^2/{\sigma^2(\beta)}}d^\ell_{mn}(\alpha)d\alpha
\eea
with the prime denoting derivative with respect to $\sigma^2(\beta)$.  Similarly, the lensed polarization power spectra can be obtained from the followings 
\begin{eqnarray}\label{lensed_clpol}
\tilde{C}_{\ell'}^{+} &=& \sum_{\ell'}  K_{\ell\ell_1}^+C_\ell^{+}, \ 
\tilde{C}_{\ell'}^{-} = \sum_{\ell'}  K_{\ell\ell_1}^-C_\ell^{-}
\end{eqnarray}
where $\tilde{C}_\cl^\pm\equiv \tilde{C}_\cl^E\pm \tilde{C}_\cl^B$, $C_\cl^\pm\equiv C_\cl^E\pm C_\cl^B$
and  $K^\pm_{\ell\ell'}$ are the associated lensing kernels for polarization given by
\bea
K^+_{\ell\ell'}&=&\delta_{\cl\cl'}+\frac{2\ell'+1}{2} \int^\pi_{0} \sin \beta \ d \beta  \  d^\ell_{22}(\beta)
\left\{\left[X^2_{022}-1\right]d^{\ell'}_{22}(\beta)+2 A_2 (\beta) X_{132}
X_{121}d^{\ell'}_{31}(\beta)\right.\nonumber\\
&+&\left.  A_2 (\beta)^2\left[X^{'2}_{022} d^{\ell'}_{22}(\beta)+X_{242}X_{220}d^{\ell'}_{40}(\beta)\right]\right\}\\
K^-_{\ell\ell'}&=& \delta_{\cl\cl'}+\frac{2\ell'+1}{2} \int^\pi_{0} \sin \beta \ d \beta  \  d^\ell_{2-2}(\beta)\left\{\left[X^2_{022}-1\right]d^{\ell'}
_{2-2}(\beta)+  A_2 (\beta) \left[X^2_{121}d^{\ell'}_{1-1}(\beta)\right.\right.\nonumber\\&+&\left.\left.
X^2_{132}d^{\ell'}_{3-3}(\beta)\right] 
+\frac{1}{2}  A_2 (\beta)^2\left[2X^{'2}_{022} d^{\ell'}_{2-2}(\beta) + X^2_{220}d^{\ell'}_{00}(\beta) +X^2_{242}d^{\ell'}_{4 -4}(\beta)\right]\right\}. 
\eea
So, given the lensing power spectrum, $C_\ell^{\phi}$, one can completely determine the lensing kernels.
\subsection{Delensed CMB power spectra}
From \eq{lensed_clt} and \eq{lensed_clpol} we see that the lensed CMB spectra are linear combinations of intrinsic  spectra 
with coefficient matrices completely determined by the lensing potential spectrum, $C_\cl^{\phi}$.  So, given the lensing potential one can, 
in principle, invert the lensing kernels and solve for the intrinsic CMB power  as shown in \citep{barun2014}. The delensed spectra obtained by direct matrix inversion can be then written as 
\bea\label{reconstructed_cmb}
C_\ell^{T}&=&\sum_{\ell_1}\left(K_{\ell\ell_1}^T\right)^{-1}\tilde{C}_{\ell_1}^{T},\\
{C_\ell^E}&=&\frac{1}{2}\bigg[C_\cl^++C_\cl^-\bigg]=\frac{1}{2}\sum_{\cl_1}\bigg[\left(K^+_{\ell\ell_1}\right)^{-1}\tilde{C}_{\cl_1}^+
+\left(K^-_{\ell\ell_1}\right)^{-1}\tilde{C}_{\cl_1}^-\bigg],\\
{C_\ell^B}&=&\frac{1}{2}\bigg[C_\cl^+-C_\cl^-\bigg]=\frac{1}{2}\sum_{\cl_1}\bigg[\left(K^+_{\ell\ell_1}\right)^{-1}\tilde{C}_{\cl_1}^+
-\left(K^-_{\ell\ell_1}\right)^{-1}\tilde{C}_{\cl_1}^-\bigg].
\eea
So given the lensing power spectrum and lensed power spectra of CMB $E$ \& $B$ modes one can reconstruct the intrinsic power for $E$ \& $B$ modes in a very simple manner.  
Also, instead of direct inversion one can make use of various numerical techniques available in the literature to solve the system of linear equations, throughout this 
work we have used {\it Gauss-Jordan} elimination technique. But here, one has be extremely careful how to deal with a truncated system of linear equations -- as $\cl$ may run from zero to infinity and we have access to very few multipoles only. We shall come back to this while discussing about delensing bias in the  followings.


\section{Delensing  in  Absence of the Instrumental Noise}\label{bias}
In our following analyses we delens the ensemble of lensed polarization spectra by solving the linear system of equations 
for four different values of tensor to scalar ratio. We have assumed, in our entire analysis, that lensing power spectrum, $C_\cl^\phi$, is completely known to us.  For  delensing we first evaluate the polarization lensing kernels $K_{\cl\cl'}^\pm$ for $\cl_{\rm max}=4096$. Then we find the inverses of the lensing kernels employing  {\it Gauss-Jordan} elimination technique and solve for the intrinsic polarization spectra.
\subsection{The Most Ideal Reconstruction}\label{ir}
\begin{figure}
	\begin{center}
		\includegraphics[width=12cm, height=8.cm]{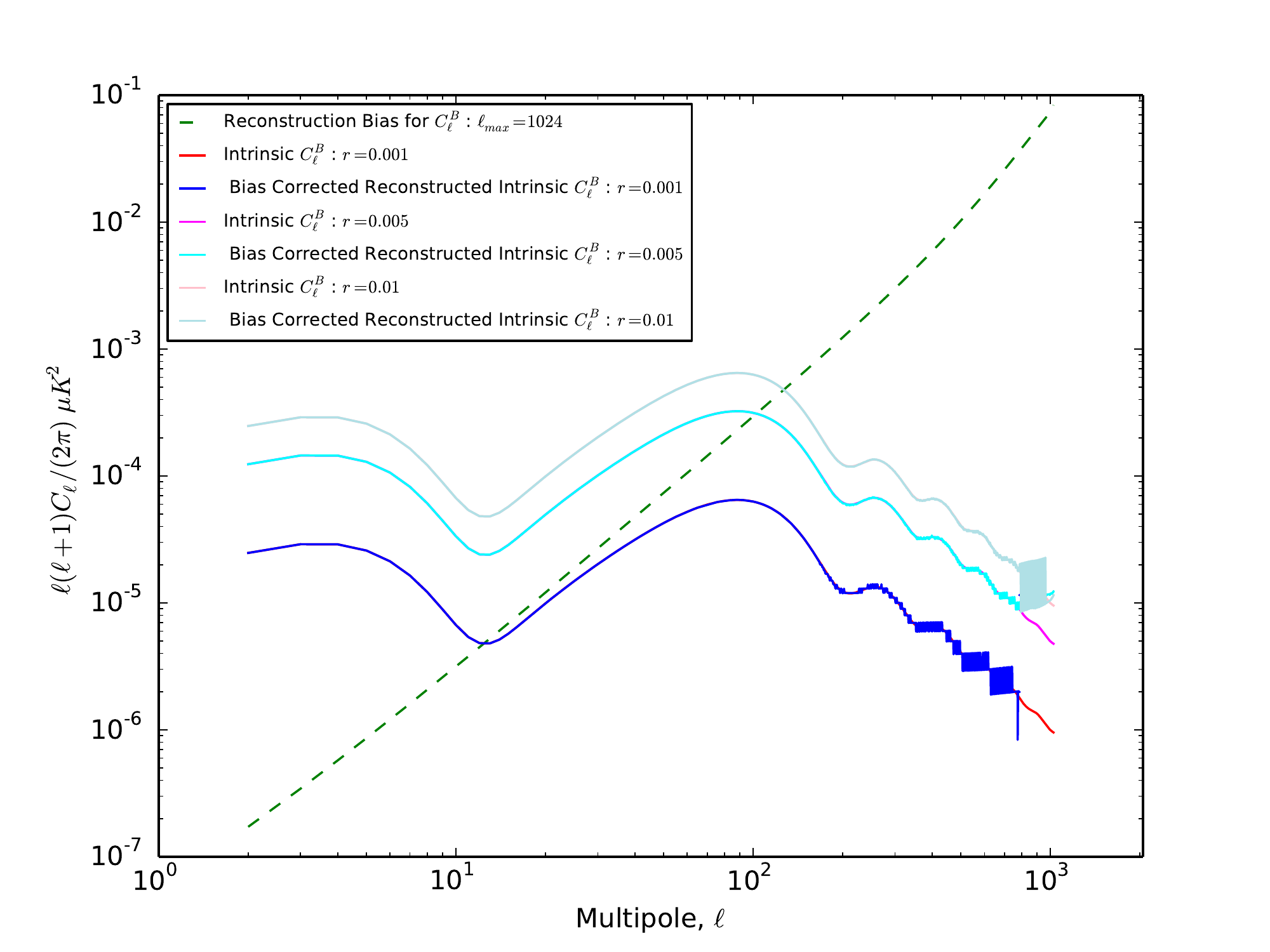}
	\end{center}
	\caption{The dashed curve shows the delensing bias for $\cl_{\rm max}=1024$. The bias subtracted reconstructed $B$ mode power along with intrinsic $C_{\ell}^B$ for three different values of tensor to scalar ratio have been plotted through the solid lines.}
	\label{test_bb}
\end{figure}

Before going into the details we demonstrate how well one can reconstruct the intrinsic $B$-mode power using our technique for a single lensed realization of CMB. In order to do so we generate the lensed $E$ and $B$ spectra for $r=0.0, \ 0.001, \ 0.005 \ {\rm and} \ 0.01$ along with the lensing potential using CAMB. We then calculate the lensing kernels and solve for intrinsic  spectra using {\it Gauss-Jordan} elimination technique. In \fig{test_bb} we have shown our results. We see that one can have a very good reconstruction of intrinsic $B$-mode power once lensing potential is known. In the \fig{test_bb} we have shown the delensing bias for $\ell_{\rm max}=1024$ which is calculated by delensing the lensed $B$ spectrum for $r=0.0$.  The reconstructed  $B$-mode power spectra are plotted after subtracting the bias for $r= 0.001, \ 0.005 \ {\rm and} \ 0.01$. So, we find that the solving the system of linear equations in order to reconstruct the intrinsic $B$ mode power might be a very useful tool. 
\subsection{Simulating Lensed Polarization Power Spectra}\label{simulation}
Thus we see that the reconstruction of large scale $B$-mode power through solving the system of linear equations may be a very useful tool when there is no error in the lensed CMB polarization spectra.  Now we want to see how efficient our technique is, when there is noise in the lensed CMB power spectra. Though we do not consider the instrumental noise for the time being, that will be discussed later in this work.  Using CAMB we generate a set of intrinsic and lensed $C_\cl^E$  \& $C_\cl^B$ for four different values of tensor to scalar ratio, $r=0, \ 0.001, \ 0.005 \ {\rm and} \ 0.01$ along with the lensing potential $C_\cl^\phi$. Since we are not considering lensing by gravity waves, $C_\cl^\phi$ is same in all those four cases. For simplicity, we generate thousand Gaussian realizations for the lensed $C_\cl^{B}$ and $C_\cl^{E}$ spectra for each value of tensor to scalar ratio setting $\cl_{\rm max}=4096$. 

\subsection{Delensing Bias -- Delensed $C_\cl^B$ in the Absence of Intrinsic $B$-mode Power}
We first apply our technique to reconstruct intrinsic $B$-mode power for $r=0$. In ideal situation one 
expects the reconstructed $C_\cl^B$ to be exactly zero. By ideal situation we mean, where we have $N$ equations in $N$ unknowns, i.e. the system is closed. But  here, in principle, one should have $N$ equations in $N$ unknowns with ${N\to\infty}$ as $\cl$ can be as large as possible. So, solving the system with finite 
$\cl_{\rm max}$ makes the solution biased which depends on what $\cl_{\rm max}$ we use for delensing, higher the $\cl_{\rm max}$ lower the bias. As a result one gets non-zero value for delensed $C_\cl^B$ even for zero tensor to scalar ratio. In order to obtain the true solution we need to subtract that bias from 
the delensed spectra we get by directly  solving the linear system. So, in the absence of instrumental noise the delensed $B-$ mode power may be read as
\beq\label{biaseq}
C_\cl^{\rm B_{del}}=C_\cl^{\rm B_{intr}}+C_\cl^{\rm B_{bias}}.
\eeq
In this case since $r=0$ we have $C_\cl^{\rm B_{intr}}=0$, therefore $C_\cl^{\rm B_{bias}}\equiv C_\cl^{\rm B_{del}}|_{r=0}$. So the above \eq{biaseq} can be written as 
\beq
C_\cl^{\rm B_{intr}}=C_\cl^{\rm B_{del}}-C_\cl^{\rm B_{bias}}. 
\eeq
Therefor the actual $B$-mode power may be obtained only after the subtraction of bias from the reconstructed $C_\cl^B$. Not only that, the actual error has to be estimated from the following equation 
\beq\label{bias_sd}
\rm{Var}(C_\cl^{\rm B_{intr}})=\rm{Var}(C_\cl^{\rm B_{del}})+\rm{Var}(C_\cl^{\rm B_{bias}})-2 \ \rm{Cov}(C_\cl^{\rm B_{del}},C_\cl^{\rm B_{bias}}). 
\eeq
\begin{figure}
	\begin{center}
		\includegraphics[width=12cm, height=8.0cm]{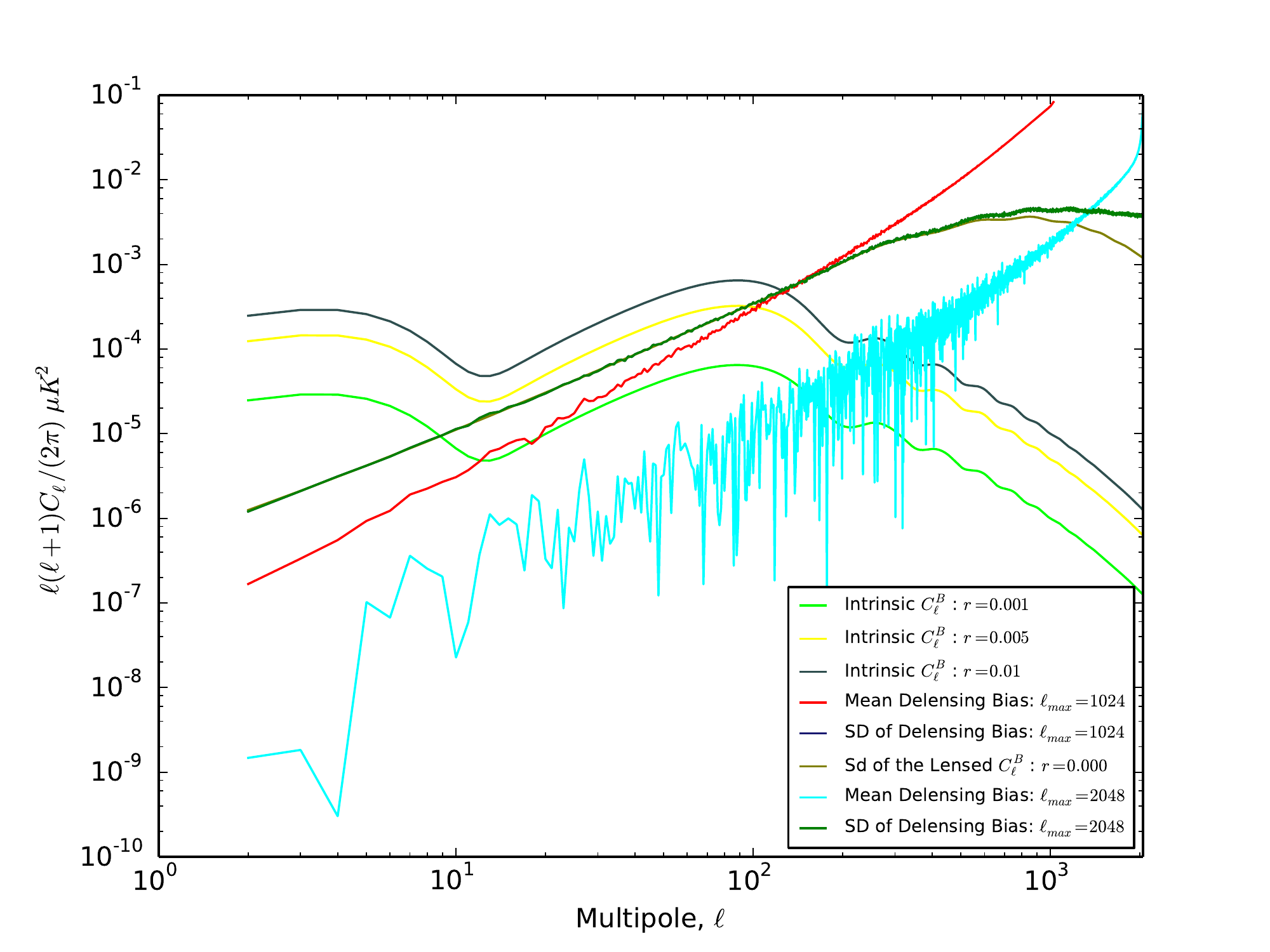}
	\end{center}
	\caption{Mean delensing biases and associated SD for $\cl_{\rm max}=1024$ and $\cl_{\rm max}=2048$. The 
		intrinsic B-mode powers for three different values of tensor to scalar ratio have also been plotted along with the SD of the lensed $C_\cl^B$ for $r=0$.}
	\label{sd_bias}
\end{figure}
In \fig{sd_bias} we have plotted mean delensing biases for $\cl_{\rm max}=1024$ and $\cl_{\rm max}=2048$ along with the
standard deviations (SD henceforth) associated with them. The mean bias is calculated from thousand delensed realizations of CMB $B$-mode power for $r=0$. The 
SDs of the biases were also estimated from those ensemble of delensed spectra. From the \fig{sd_bias} we also see that mean bias is  smaller when we use higher $\cl_{\rm max}$, but corresponding SDs are almost same. Here we also note that the SDs associated with the delensing biases are almost same as that of the lensed $C_\cl^B$ for $r=0$.    
\subsection{Delensed $C_\cl^B$ for Three Different Values of Tensor to Scalar Ratio}
\begin{figure}
	\begin{center}
		\includegraphics[width=8.8cm, height=8.cm]{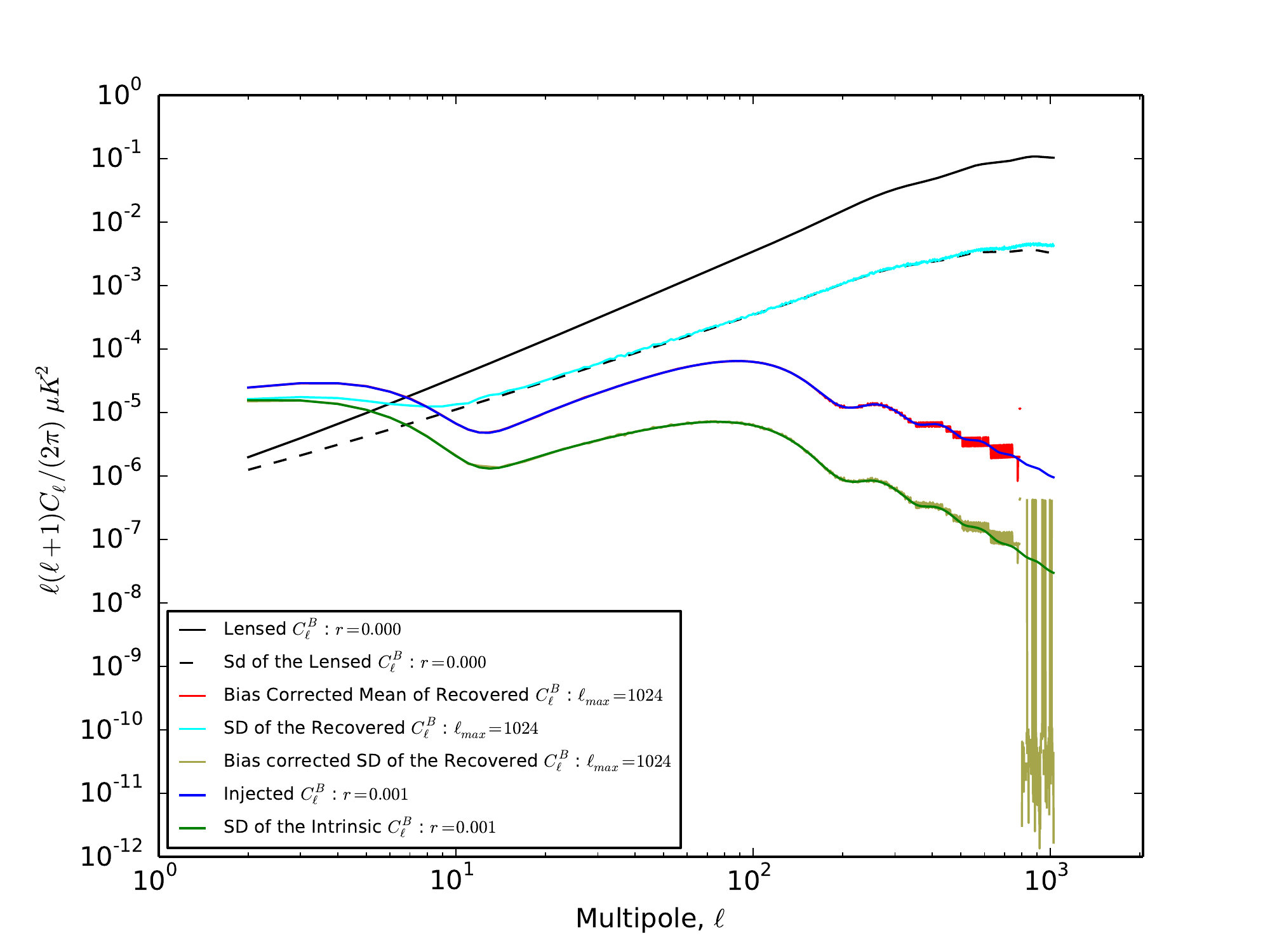}
		\includegraphics[width=8.8cm, height=8.cm]{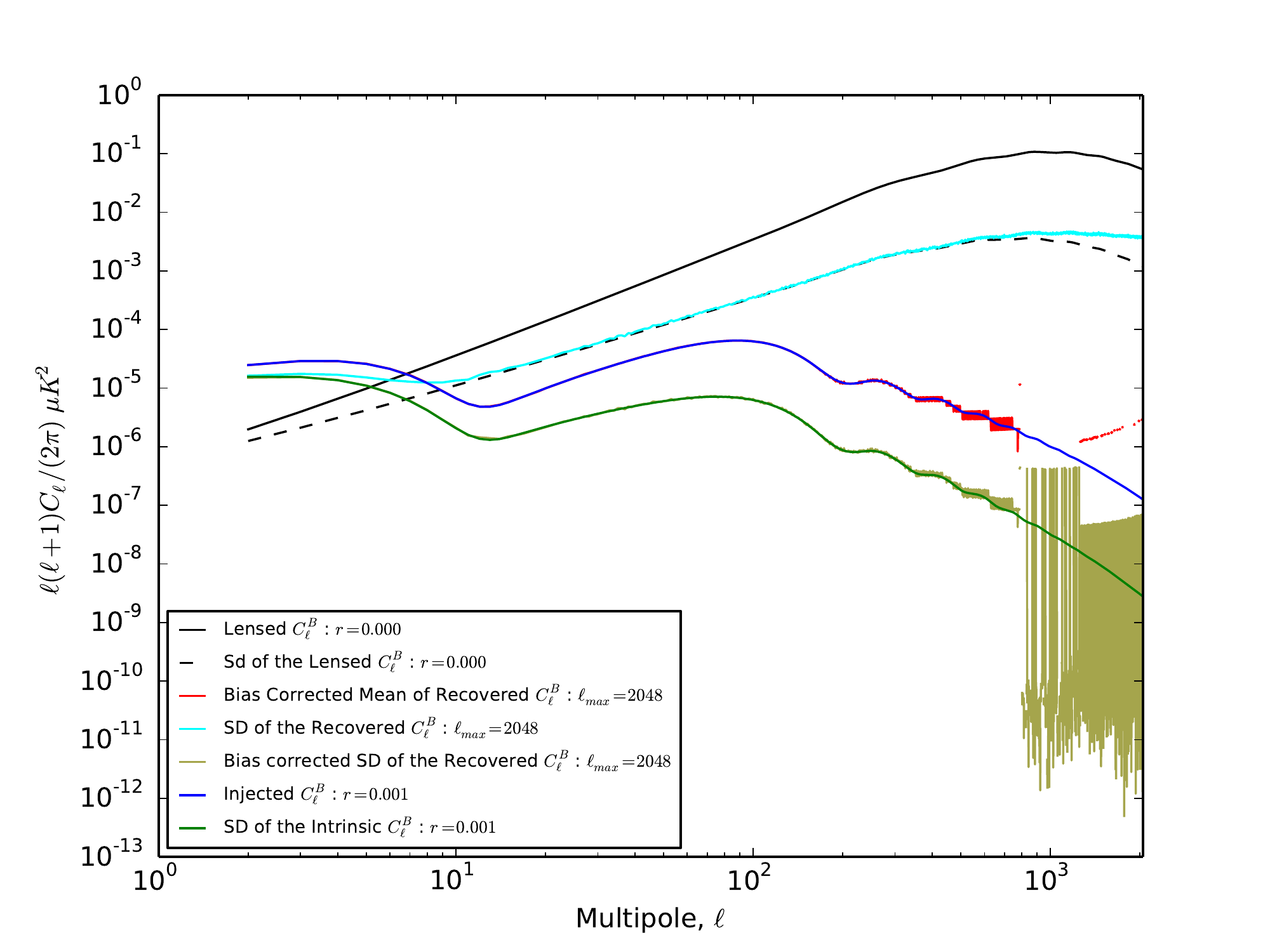}
	\end{center}
	\caption{Delensed $C_\cl^B$ after subtracting the bias for $r=0.001$, $\cl_{\rm max}=1024$ and the SD. Corresponding intrinsic $C_\cl^B$, mean 
		delensing bias, SD of the bias and the SD of the lensed $C_\cl^B$ for $r=0$ have been plotted. In the right panel the result for $\cl_{\rm max}=2048$
		has also been plotted.}
	\label{sim_001}
\end{figure}

We now employ our delensing algorithm to the lensed realizations for  $r=0.001$  with $\cl_{\rm max}=1024$. From the ensemble of delensed spectra we calculate the mean and subtract the mean delensing bias from it to get the intrinsic $C_\cl^{\rm B_{intr}}$. In \fig{sim_001} we have plotted the mean of the recovered $C_\cl^B$ after subtracting the mean delensing bias along with the SD and injected $C_\cl^B$. We can see that bias subtracted  mean of the recovered $C_\cl^{B}$ matches very well with the injected $C_\cl^{B}$.  Now \fig{sim_001} also reveals that the reconstruction noise is too high to distinguish the intrinsic $B-$ mode signal from the  noise. But if we use \eq{bias_sd} to estimate the SD for the reconstructed intrinsic  $C_\cl^{B}$, one can have CVL  reconstruction of the intrinsic $C_\cl^B$ up to few hundred multipoles in absence of instrumental noise.  In the right panel of \fig{sim_001} we have shown the results for $\cl_{\rm max}=2048$. It is also obvious from the figure that employment of higher $\cl_{\rm max}=2048$ for delensing does not have much impact in our reconstruction process. 

\begin{figure}
	\begin{center}
		\includegraphics[width=8.8cm, height=8.cm]{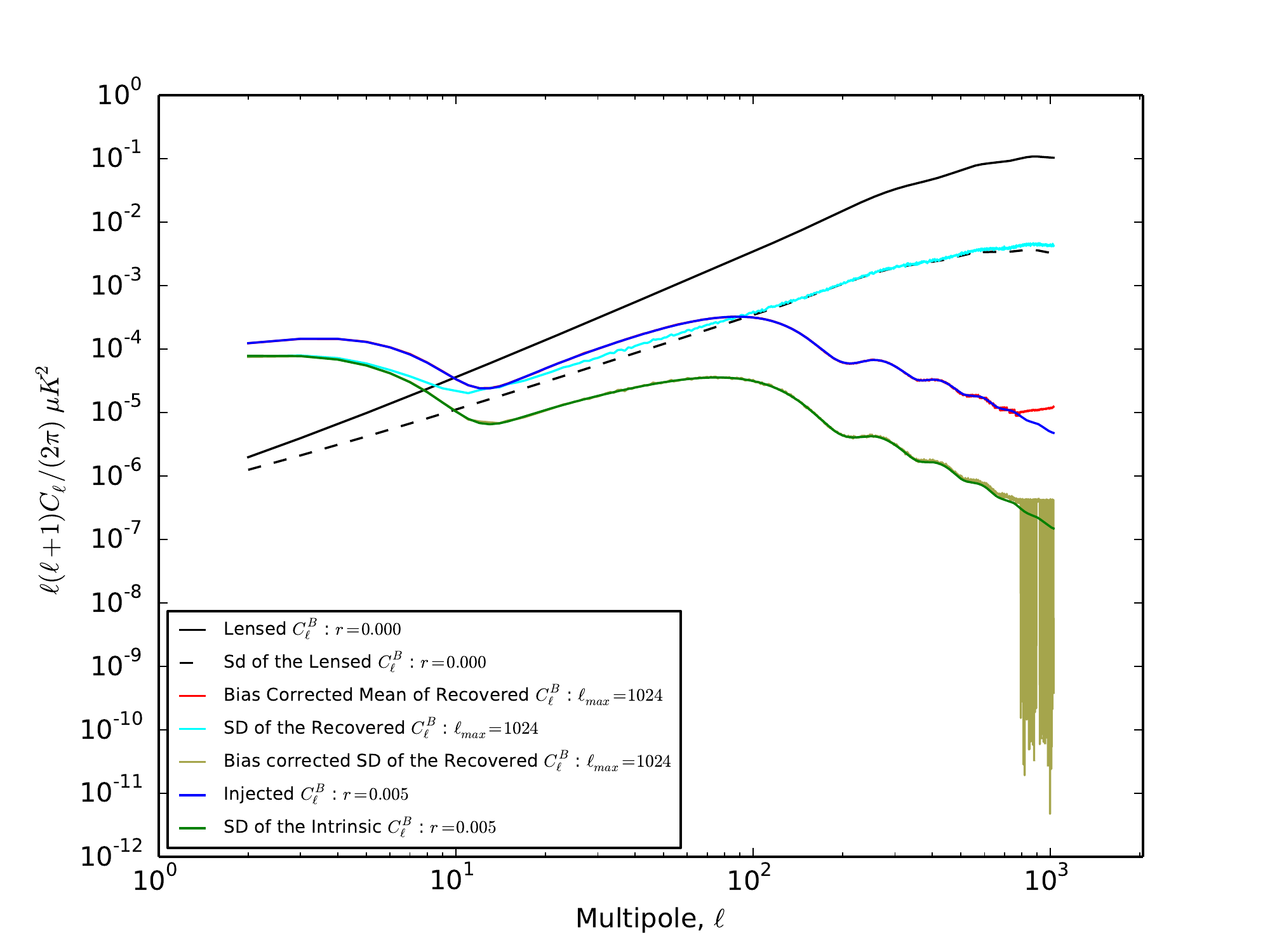}
		\includegraphics[width=8.8cm, height=8.cm]{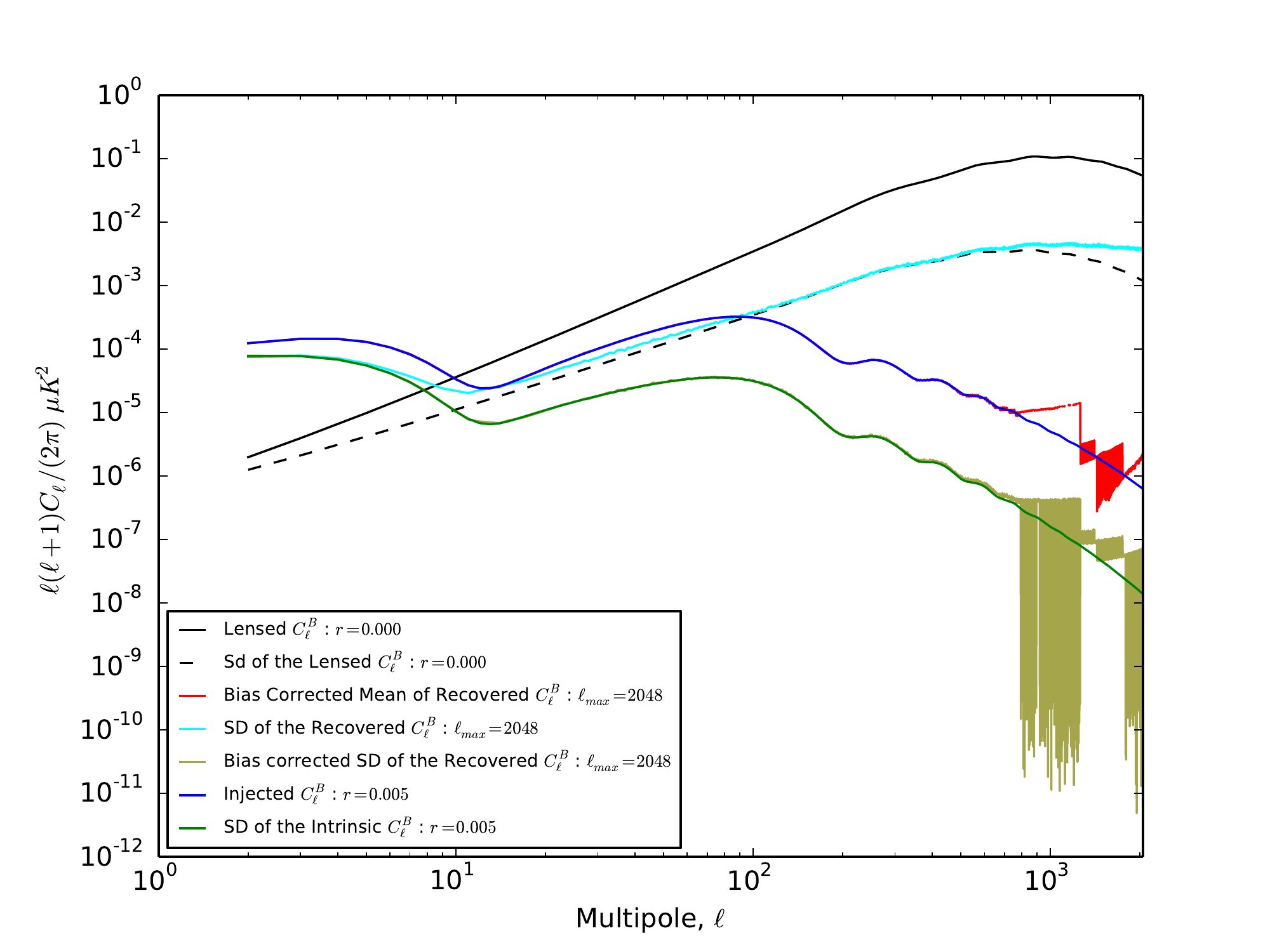}
	\end{center}
	\caption{Delensed $C_\cl^B$ after subtracting the bias for $r=0.005$, $\cl_{\rm max}=1024$ and the SD. Corresponding intrinsic $C_\cl^B$, mean 
		delensing bias, SD of the bias and the SD of the lensed $C_\cl^B$ for $r=0$ have been plotted. In the right panel the same with $\cl_{\rm max}=2048$ has 
		been shown.}
	\label{sim_005}
\end{figure}

Next we repeat the above procedure to delens the lensed realizations for $r=0.005$ with $\cl_{\rm max}=1024$. We again subtract the mean delensing bias from the 
mean of the delensed spectra to get the intrinsic $C_\cl^{\rm B_{intr}}$. In \fig{sim_005} 
we have plotted the bias subtracted mean of the recovered $C_\cl^B$ along with the SD and injected $C_\cl^B$. From \fig{sim_005} we see that one can have CVL reconstruction of the intrinsic $C_\cl^B$ up to few hundred multipoles. Also, from the right panel of \fig{sim_005} we conclude that using higher delensing $\cl_{\rm max}$ does not provide any different result. 

\begin{figure}
	\begin{center}
		\includegraphics[width=8.8cm, height=8.cm]{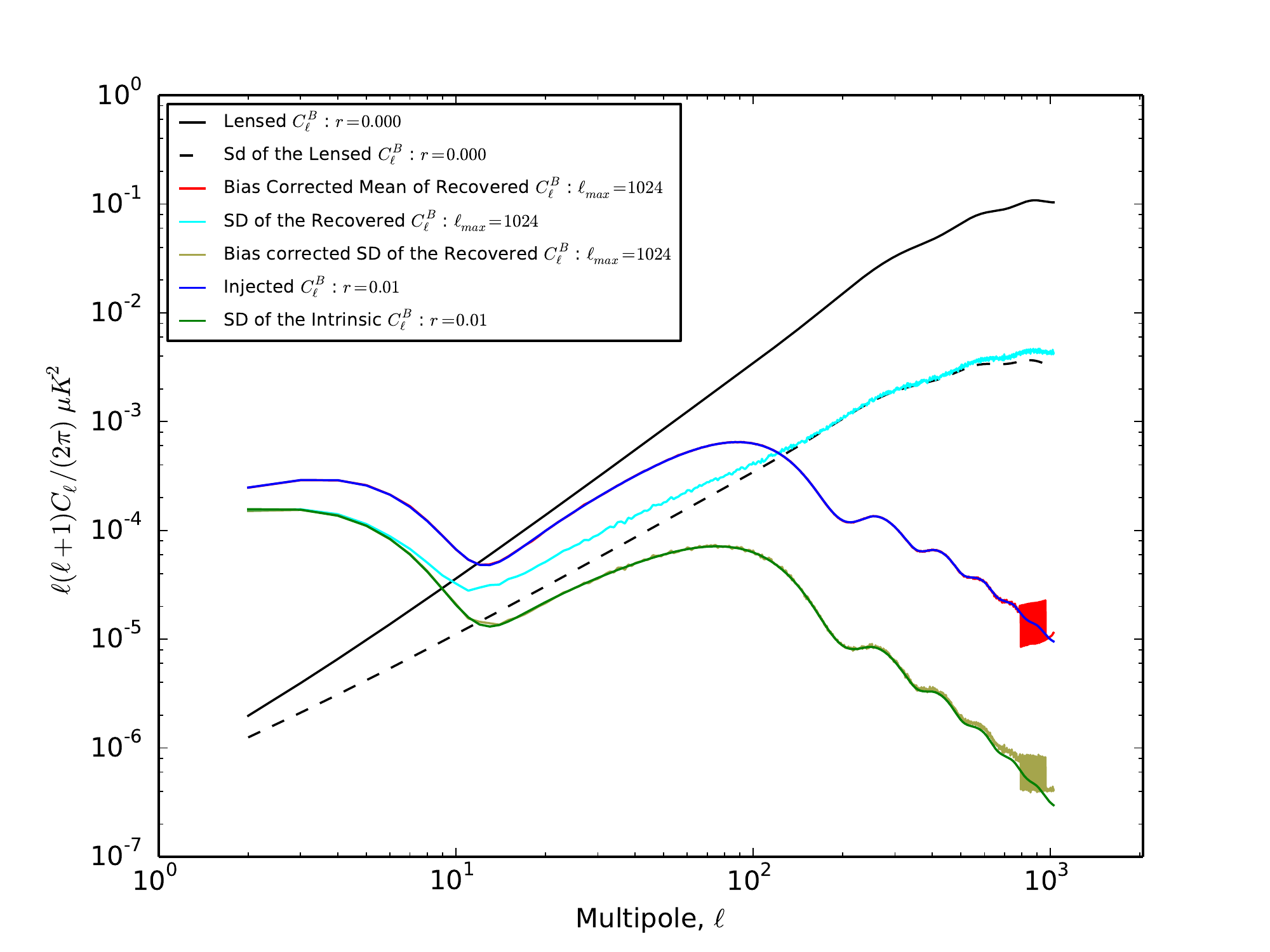}
		\includegraphics[width=8.8cm, height=8.cm]{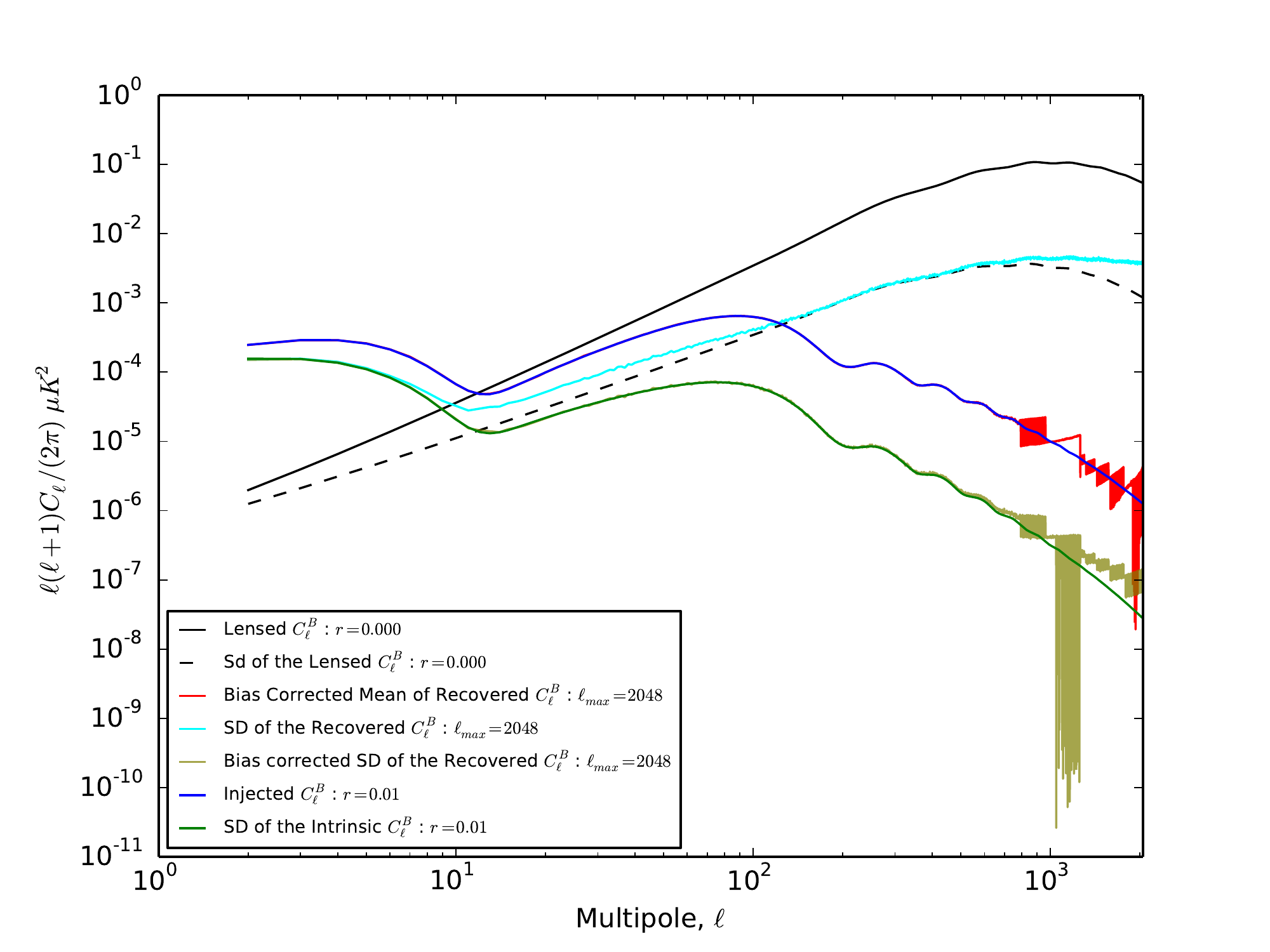}
	\end{center}
	\caption{Delensed $C_\cl^B$ after subtracting the bias for $r=0.01$, $\cl_{\rm max}=1024$ and the SD. Corresponding intrinsic $C_\cl^B$, mean 
		delensing bias, SD of the bias and the SD of the lensed $C_\cl^B$ for $r=0$ have been plotted. In the right panel the same with $\cl_{\rm max}=2048$ has 
		been plotted.}
	\label{sim_010}
\end{figure}

Finally we employ our technique for the case $r=0.01$ and delensing $\cl_{\rm max}=1024$. From the solution set for delensed spectra we calculate the mean and subtract the mean delensing bias  to get the intrinsic $C_\cl^{\rm B_{intr}}$. As before to get the actual SD we have used the \eq{bias_sd}.   
In \fig{sim_010} we have plotted the bias subtracted mean of the recovered 
$C_\cl^B$ along with the bias corrected SD and injected $C_\cl^B$. From \fig{sim_010} it is clear that one can have CVL reconstruction of the intrinsic $C_\cl^B$ beyond $\ell\sim 500$. The right panel of \fig{sim_010} we have demonstrated the result with $\cl_{\rm max}=2048$.

Thus,  our above analysis reveals that  solving the linear system to get the intrinsic $C_\cl^B$ is indeed a very powerful method, provided we have the cosmological parameters in advance specially the lensing power spectrum and lensed polarization spectra. We find that using higher $\cl_{\rm max}$  does not provide any better result and employment of $\cl_{\rm max}=1024$ for delensing is more than sufficient in oder to unveil the primordial gravity waves. We see that in the absence of instrumental noise it is possible to have CVL reconstruction of intrinsic $B$-mode power up to $\cl\sim500$ even for $r=0.001$ which is indeed fascinating. So this reconstruction algorithm my be applied for a wide range of values of tensor to scalar ratio in any noise free experiment. 
\section{Delensing in Presence of the Instrumental Noise}\label{del_noise}
In this section we repeat the above exercise incorporating instrumental noise anticipating  COrE like futuristic space mission. For the rest of the work we have considered only delensing $\cl_{\rm max}=1024$. The noise power spectrum is given by \citep{knox1995, tegmark1997, perotto2006, galli2014} 
\beq\label{noise}
N_\cl = \sigma^2\theta_{\rm fwhm}^2\ \exp{\left[\cl(\cl+1)\frac{\theta_{\rm fwhm}^2}{8\ln 2}\right]} 
\eeq
where $\theta_{\rm fwhm}$ is the {full width half maximum} (in radian) of the Gaussian beam and $\sigma$ is the root mean square of the 
detector noise for polarization having unit $\mu K$. 
For the estimation of the noise power spectrum we have considered COrE's $225$ GHz channel with specifications as follows: $\theta_{\rm fwhm}=4.7$ arcmin,  
$\sigma=4.57$ $\mu K$-arcmin for polarization and number of detectors $=1800$. We first evaluate the noise power spectrum using $\cl_{\rm max}=4096$ for COrE's $225$ GHz frequency channel and then generate 1000 realizations of the noise spectra. 

The observed $B$-mode power in presence of the instrumental noise may be written as
\bea
C_\cl^{\rm B_{\rm obs}}&=&C_\cl^{\rm B_{\rm intr}}+C_\cl^{\rm B_{\rm len}}+N_\cl^{\rm B_{noise}}.
\eea
When the observed spectrum is delensed that $C_\cl^{B_{\rm del}}$ will also contain, apart from the intrinsic signal, the contribution from the instrumental noise as well as the bias. Therefore the delensed $B$-mode power may now be written as 
\bea C_\cl^{\rm B_{\rm del}}&=&C_\cl^{\rm B_{\rm intr}}+C_\cl^{\rm B_{\rm bias}}+N_\cl^{\rm B_{noise}}.
\eea
So in order to get the true $B$-mode power we now need to subtract the delensing bias along with instrumental noise from  $C_\cl^{\rm B_{\rm del}}$. The detail procedure has been discussed in the following. 
\subsection{Delensing Bias in Presence of Instrumental Noise}
As we delens the observed $C_\cl^{\rm B_{\rm obs}}$ with $r=0$ in presence of non-zero instrumental noise, the delensed spectra is again non-zero. So the delensing bias in presence of instrumental noise may be obtained by delensing the lensed spectra (after adding instrumental noise, assuming noise and signal to be uncorrelated) for $r=0$. As a result the intrinsic $B$ mode power may be found after the subtraction of bias and instrumental noise from the delensed $B$ mode power i.e., 
\bea
C_\cl^{\rm B_{\rm intr}}&=&C_\cl^{\rm B_{\rm del}} -\left(C_\cl^{\rm B_{\rm bias}}+N_\cl^{\rm B_{noise}}\right)\nonumber\\
&=&C_\cl^{\rm B_{\rm del}} -{C_\cl^{\rm B_{\rm BN}}}
\eea 
where we have defined $C_\cl^{\rm B_{\rm BN}}\equiv C_\cl^{\rm B_{\rm bias}}+N_\cl^{\rm B_{noise}}=C_\cl^{\rm B_{\rm del}}|_{r=0}$ as the delensing bias in presence of instrumental noise. So, the actual error in the reconstructed $B$-mode power may be written as 
\beq\label{bias_sd_noise}
{\rm Var}(C_\cl^{\rm B_{\rm intr}})={\rm Var}(C_\cl^{\rm B_{\rm del}})+{\rm Var}(C_\cl^{\rm B_{\rm BN}}) -2 \ {\rm Cov}\left(C_\cl^{\rm B_{\rm del}},C_\cl^{\rm B_{\rm BN}}\right). 
\eeq
In estimating the bias after taking into account instrumental noise we may be able to overcome limitation imposed by our instrument. In Fig.\ref{bias_noise} we have presented the delensing bias in presence of instrumental noise, $C_\cl^{\rm B_{\rm BN}}$.
The mean delensing bias and the SD were estimated from the 1000 delensed $C_\cl^{\rm B}$ spectra for zero tensor to scalar ratio assuming COrE like mission. 
\subsection{Delensed $C_\cl^B$ for $r=0.001$, $r=0.005$ and $r=0.01$ with COrE Like Instrumental Noise}
\begin{figure}
	\centering
	\centering
	\begin{minipage}{0.4\linewidth}
		\includegraphics[width=8.cm, height=7.5cm]{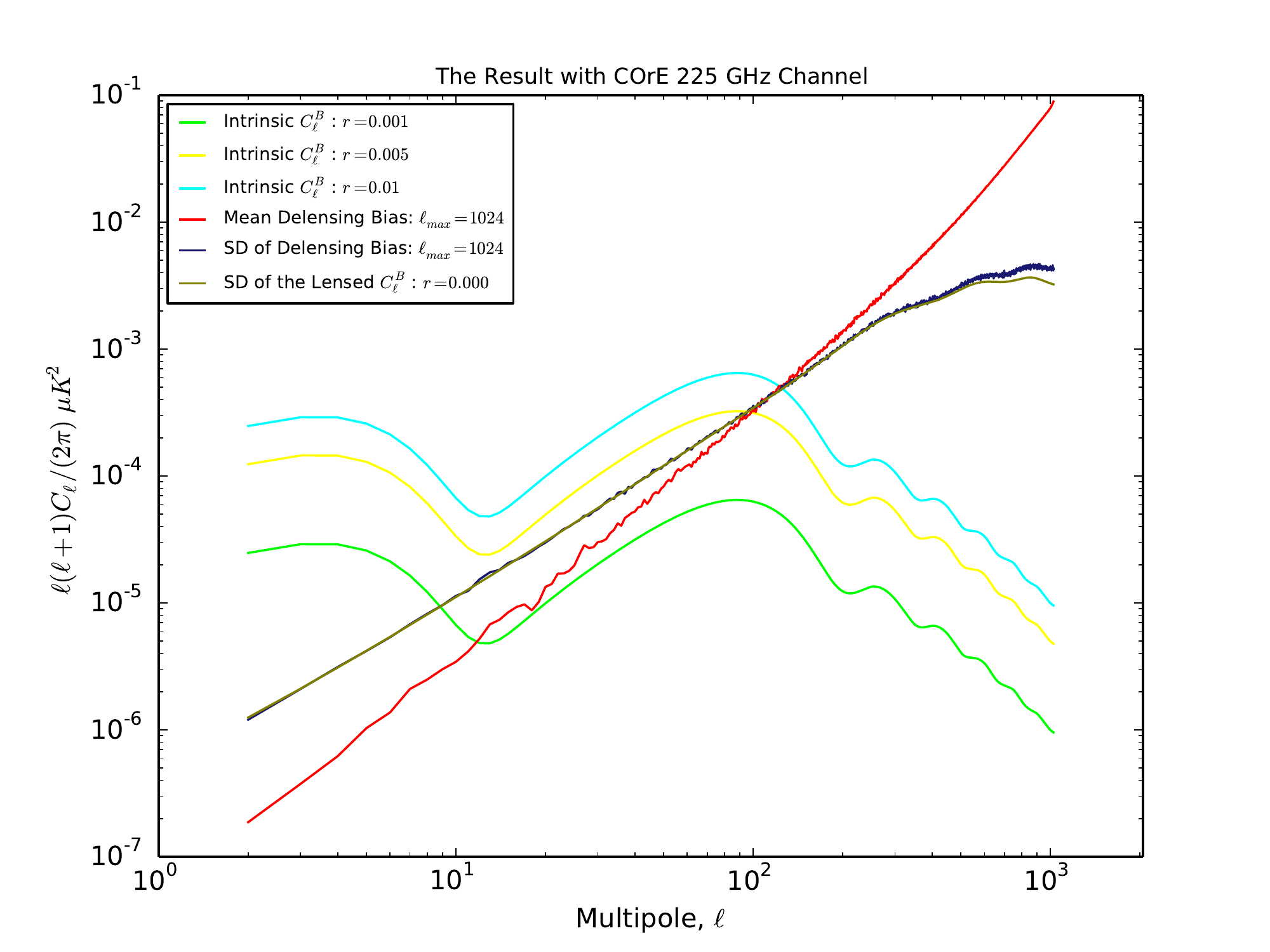}
		\caption{Plot of mean delensing bias in presence of instrumental noise. Intrinsic $C_\ell^B$ for $r=0.001, \ 0.005 \ \mbox{and} \ 0.01$ along with the mean delensing bias with $\ell_{\rm max}=1024$ have been plotted. The SD of the delensing bias has also been plotted.}
		\label{bias_noise}
	\end{minipage}
	\quad
	\centering
	\begin{minipage}{0.4\linewidth}
		\includegraphics[width=8.cm, height=7.5cm]{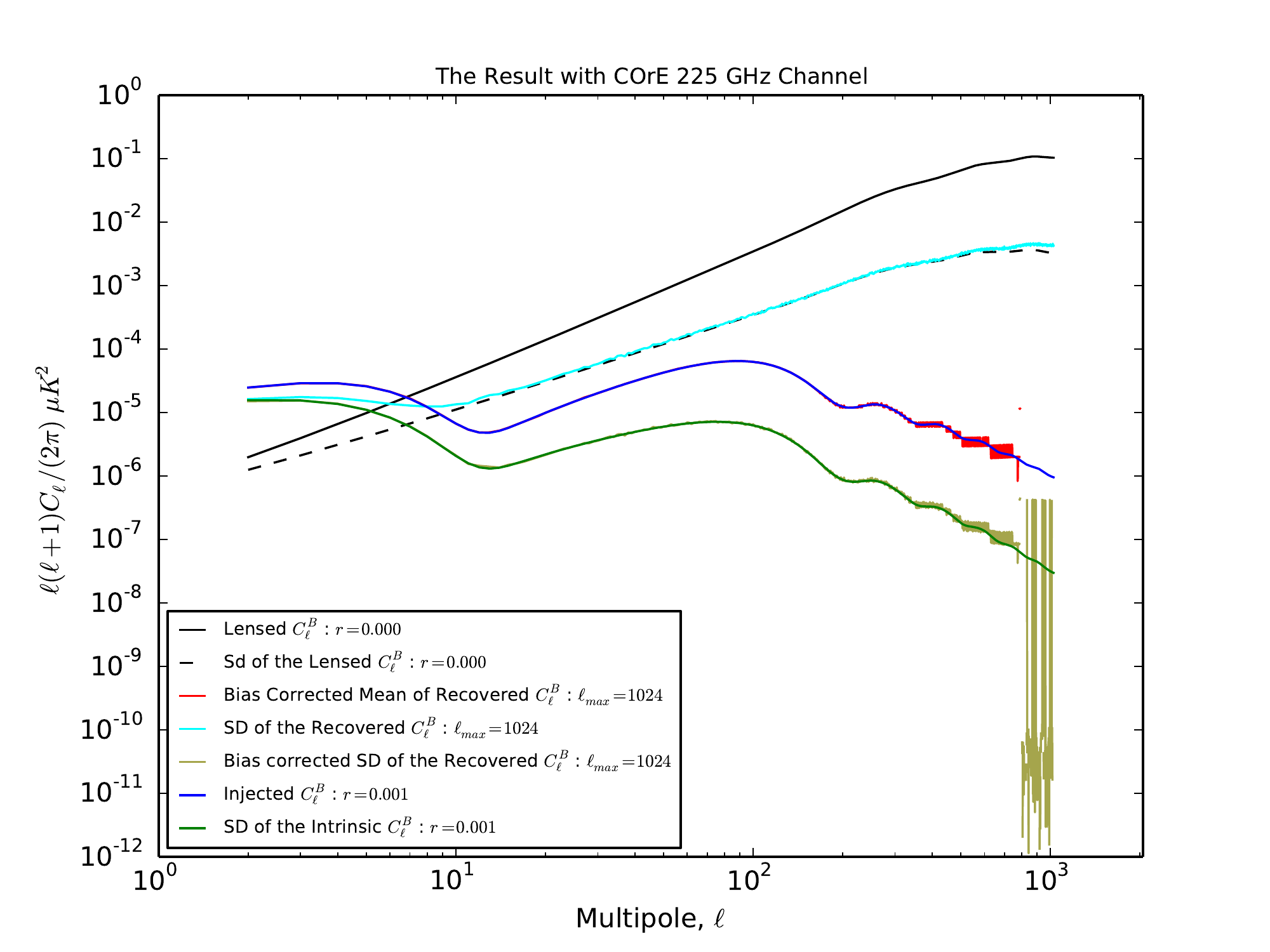}
		\caption{Mean of the delensed $C_\cl^B$ after subtracting the bias for $r=0.001$, $\cl_{\rm max}=1024$ and the SD have been plotted. Corresponding intrinsic $C_\cl^B$ and SD for $r=0.001$ along with bias corrected SD of the reconstructed $C_\cl^B$ have been plotted.}
		\label{noise_001}
	\end{minipage}
	\quad
	\begin{minipage}{0.4\linewidth}
		\includegraphics[width=8.cm, height=7.5cm]{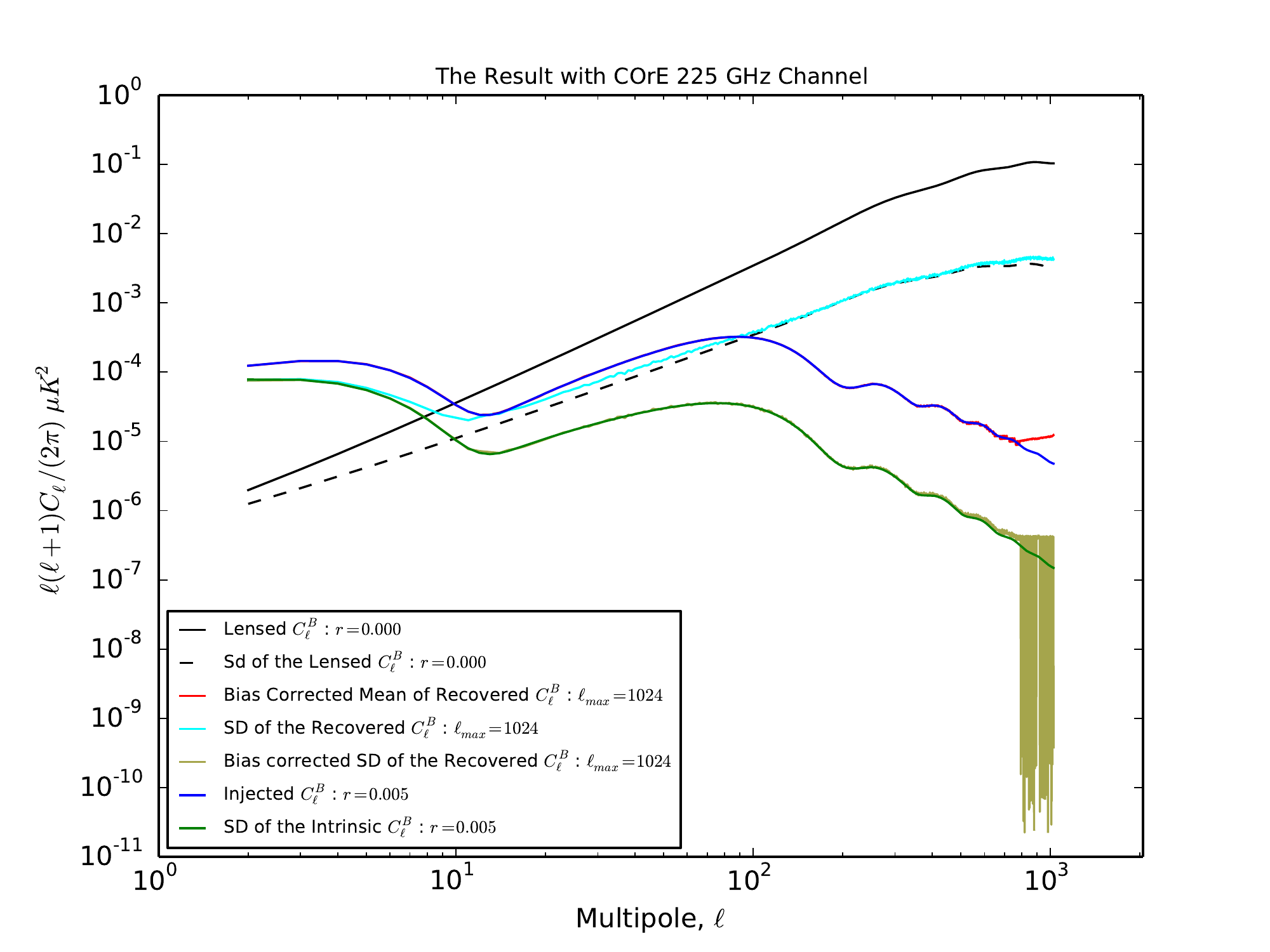}
		\caption{Bias subtracted mean of the delensed $C_\cl^B$ for $r=0.005$ with $\cl_{\rm max}=1024$ and corresponding SD have been shown. The intrinsic $C_\cl^B$ and its SD along with bias corrected SD of the reconstructed $C_\cl^B$ also have been plotted.}
		\label{noise_005}
	\end{minipage}
	\quad
	\begin{minipage}{0.4\linewidth}
		\includegraphics[width=8.cm, height=7.5cm]{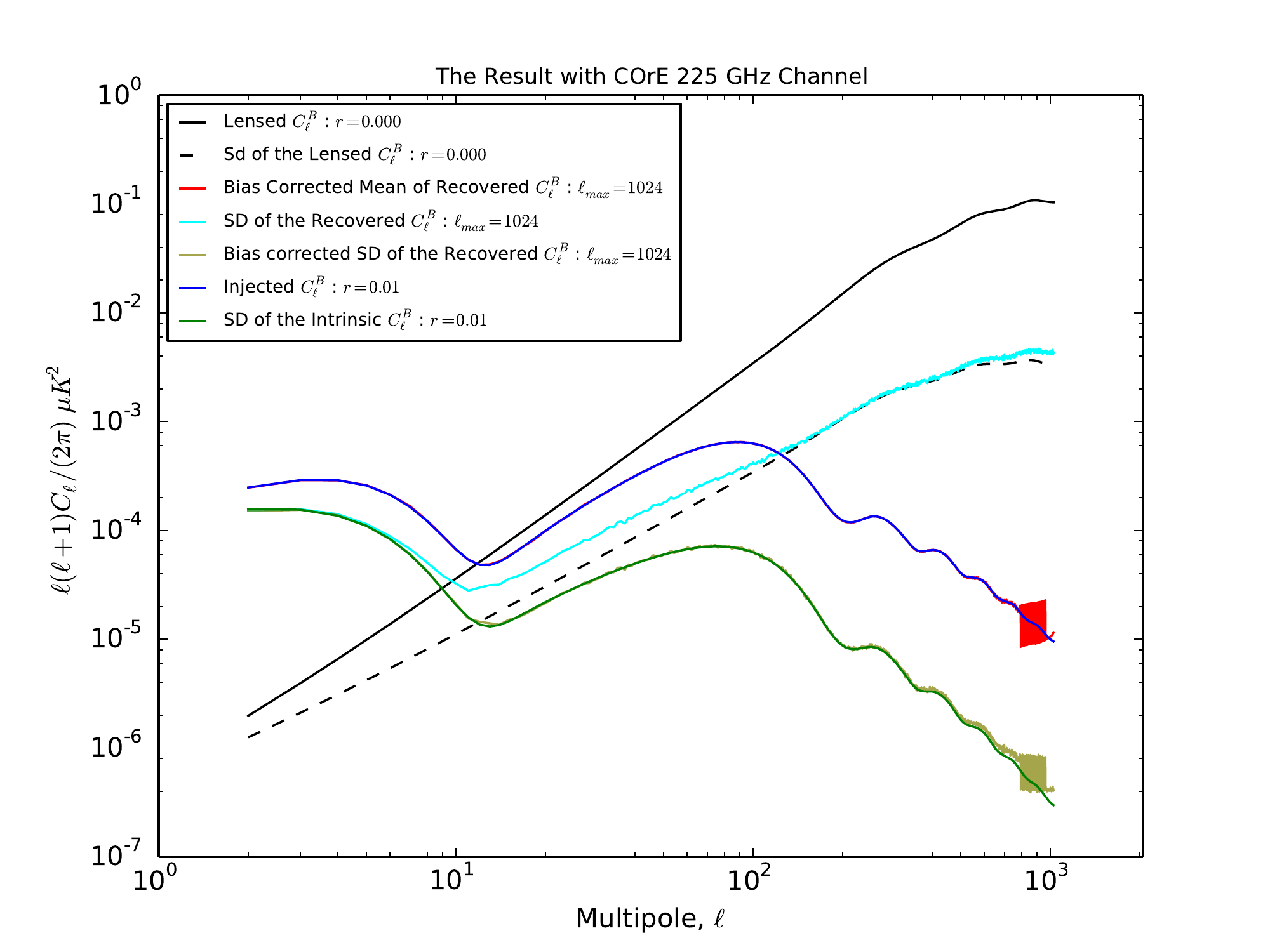}
		\caption{Bias subtracted mean of the delensed $C_\cl^B$ for $r=0.01$ with $\cl_{\rm max}=1024$ and corresponding SD have been shown. The intrinsic $C_\cl^B$ and its SD along with bias corrected SD of the reconstructed $C_\cl^B$ also have been plotted.}
		\label{noise_010}
	\end{minipage}
\end{figure}

Finally we employ our delensing algorithm to the lensed realizations  anticipating COrE like futuristic satellite mission for non-zero tensor to scalar ratio with $\cl_{\rm max}=1024$. We first add the noise to the lensed CMB $B$  and $E$  realizations separately. Then we delens those noisy lensed realizations and analyze the result.  In order to get hold of intrinsic $C_\cl^{\rm B_{\rm intr}}$ we subtract the mean delensing bias from the mean of the delensed $C_\cl^B$. The actual error of intrinsic $C_\cl^{\rm B_{\rm intr}}$ has been estimated using \eq{bias_sd_noise}. The results for $r=0.001, \ 0.005 \ {\rm and} \ 0.01$ have been shown in the Figs.\ref{noise_001}, \ref{noise_005} and \ref{noise_010} respectively. From those figures we find that the bias corrected mean and SDs of the recovered $B$-mode power are in good agreement with those of the intrinsic $C_\cl^{\rm B}$ for all the three values of tensor to scalar ratio. Also, since we have calculated the bias after adding the noise with lensed polarization spectra, the impact of instrumental noise does not play any significant role in our reconstruction. 

\begin{figure}
	\begin{center}
		\includegraphics[width=8.8cm, height=8.cm]{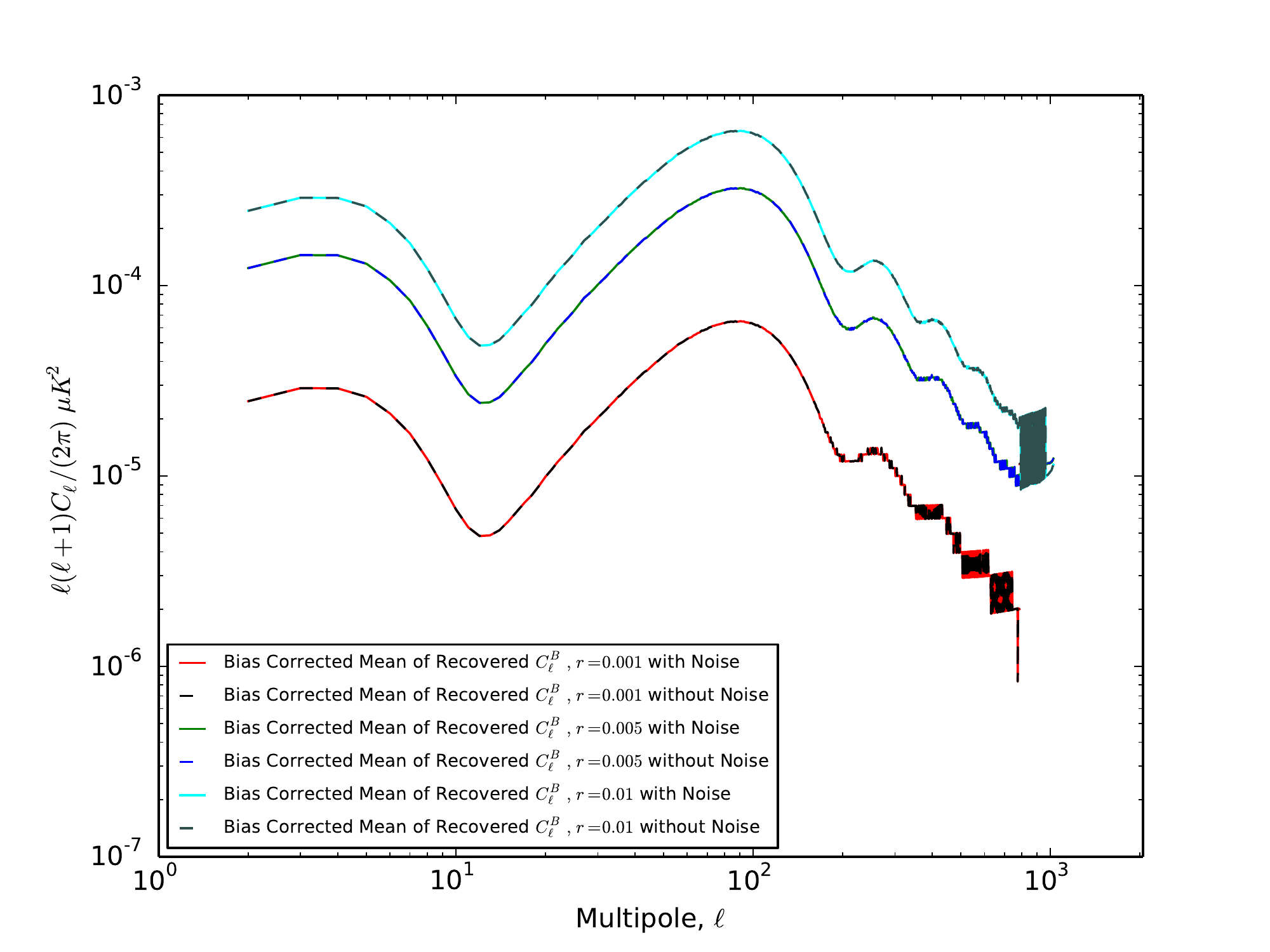}
		\includegraphics[width=8.8cm, height=8.cm]{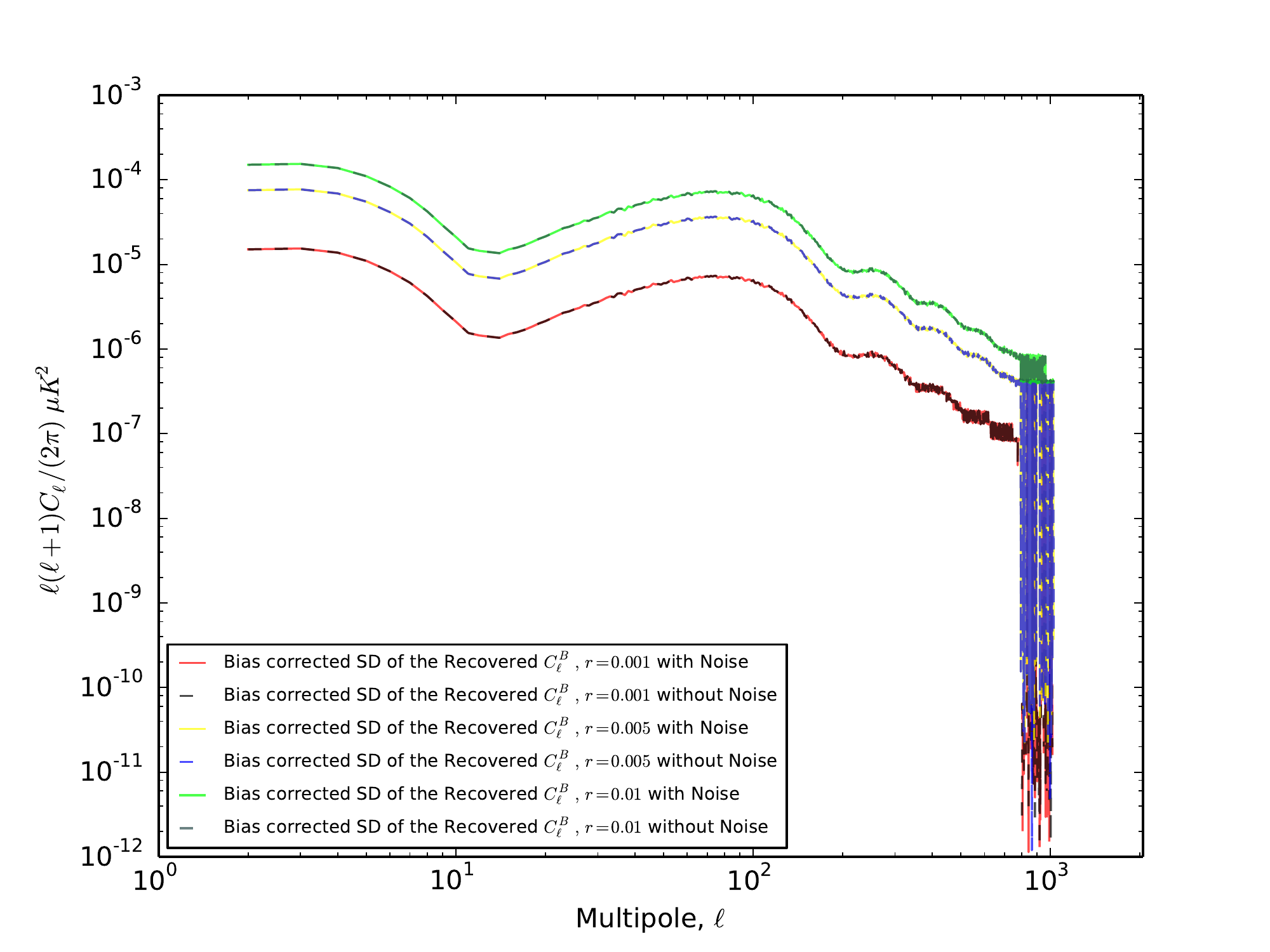}
	\end{center}
	\caption{Left Panel: A comparison between bias corrected means with and without taking into account the instrumental noise for three different values of the tensor to scalar ratio have been plotted.  
		In the right panel the same but now for the bias corrected SDs for three different values of $r$ have been shown.}
	\label{comparison}
\end{figure}
The comparison between results for zero and non-zero instrumental noises have been shown in the \fig{comparison}  for three different values of tensor to scalar ratio with delensing $\cl_{\rm max}=1024$. We find that the bias corrected mean and SD are almost same whether we consider the instrumental noise or not. The actual  explanation is quite easy to understand. We are calculating the bias after adding noise to the lensed CMB. Consequently, the impact of noise gets hidden into the bias itself. As a result, when we subtract the mean delensing bias from that of the delensed realization the effect of instrumental noise is cancelled and we have similar looking plots in \fig{comparison}. Therefore we may assert that the impact of the instrumental noise may be bypassed within our algorithm. Not only that this method may applied to reconstruct the intrinsic $B$-mode power for a broad  range of tensor to scalar ratio. 
\section{Conclusion}
In this article we have demonstrated a simple algorithm for the spectral level reconstruction of CMB $B$-mode power. The algorithm consists of solving the simultaneous system of linear equations by finding the inverses of lensing kernels using {\it Gauss-Jordan} elimination technique. We find that one can have CVL reconstruction of intrinsic $B$-mode power up to few hundred multipoles which is more than the range we are searching for the  primordial gravity waves, even for $r$ as low as $10^{-3}$. Though our current observations are far away to make a detection of $r=0.001$, but the futuristic space missions in the likes of  PRISM, COrE may be able to detect signal of primordial gravity waves with tensor to scalar ratio as low as $10^{-3}$ or even a order lower. Accordingly our methodology may be applied to those futuristic space missions as well. We also find that the incorporation of COrE like instrumental noise does not affect our reconstruction. Actually, the impact of instrumental noise may be bypassed within our algorithm. Though we do not make any hefty comments on this, as the non-Gaussian structure of the lensed CMB has been neglected along with the lensing by gravity waves, but still we are optimistic. The errors on the cosmological parameters has not been considered as well. But still  one may assert that this method of spectral level reconstruction has the potential to serve our purposes in uncovering the primordial gravity waves.  

The lensing induced by gravity waves may not be significant as it is expected to be very small. So it can be safely neglected. We also do not consider the fact that the observations only cover around $70-80\%$ of the sky. And that partial sky coverage will bring in more noise in the reconstructed intrinsic spectrum. A more rigorous analysis can also be done by taking into account the non-zero correlation between lensed $E$ and $B$-modes which seeks simulation of lensed CMB sky. But at a first go realizing the simplicity of the method, the assumptions made therein are justified  for the time being. We hope to comeback with more sophisticated analysis incorporating the non-Gaussian structure of the lensed CMB along with the effect of  partial sky coverage in near future.     
\section*{Acknowledgements}
Author is very grateful to Aditya Rotti and Supratik Pal for useful discussions and suggestions. Author also acknowledges the use of publicly available code CAMB for his analysis.



\bibliographystyle{mnras}
\bibliography{references} 


\bsp	
\label{lastpage}
\end{document}